\documentclass[12pt,letterpaper]{article}
\usepackage[utf8]{inputenc}
\usepackage{amsmath}
\usepackage{amsfonts}
\usepackage{amssymb}
\usepackage{cite}

\usepackage{xcolor}

\usepackage[hidelinks]{hyperref}

\usepackage[letterpaper,left=2cm, right=2cm, top=3cm, bottom=3cm]{geometry} 

\numberwithin{equation}{section}

\title{Palatini Gauss-Bonnet theory}

\begin{document}

\author{M\'aximo Ba\~nados$^1$ and Marc Henneaux$^2$  \\
{\footnotesize $^1$ Facultad de F\'isica, Pontificia Universidad Cat\'olica de Chile,  Santiago, Chile}  \and   
{\footnotesize $^{ 2}$ Universit\'e Libre de Bruxelles and International Solvay Institutes,
Brussels, Belgium} \\ {\footnotesize and Coll\`ege de France,  Universit\'e PSL,   Paris, France}}

\maketitle 

\abstract{We consider a class of models in even spacetime dimensions $2n$ which share many similarities with Chern-Simons theories in odd spacetime dimensions $2n+1$.  The independent dynamical variables of these models are a $GL(2n)$-connection and a metric in internal space.  The action is a polynomial of degree $n$ in the curvature of the connection, with indices saturated by means of the metric and the Levi-Civita tensor.  We show that the theory has no local degree of freedom in $2$ spacetime dimensions ($n=1$), where it can be reformulated as a constrained $BF$ model, but that its dynamics is more intrincate in higher dimensions ($n>1$), where local degrees of freedom are present.  We treat in detail the cases of $2$ and $4$ spacetime dimensions.}

\section{Introduction}

Pure Chern-Simons theory in $3$ spacetime dimensions has been a constant source of inspiration since its inception in the 1970-1980's \cite{Schwarz:1978cn,Witten:1988hf}.  Chern-Simons theories exist also in higher dimensions, but only in the odd-dimensional case.  

It is the purpose of this paper to study a class of theories, named ``Palatini Gauss-Bonnet theories", which exist only in even spacetime dimensions $2n$ and share many of the features of the Chern-Simons models in dimensions $2n+1$: (i)  These theories are topological in the sense that they do not involve a spacetime metric. (ii) The coupling constant is dimensionless. (iii) In $2$ spacetime dimensions, the Lagrangian is non-trivial (not equal to a total divergence) but  
there is no local degree of freedom due to the big number of gauge symmetries. (iv) The theory does possess local degrees of freedom when $n>1$, but identifying them turns out to be an intricate question, as for pure Chern-Simons theory in $2n+1$ dimensions with $n\geq 2$ \cite{Banados:1995mq,Banados:1996yj}.

\subsection{Action}

The dynamical variables of the Palatini Gauss-Bonnet models are a $GL(2n)$-connection $1$-form ${\Gamma^\alpha}_{\beta} \equiv {\Gamma^\alpha}_{\beta \mu}dx^\mu$ and a metric $g_{\alpha \beta}$ in internal space \footnote{Internal indices will be taken from the first half of the Greek alphabet,  $\alpha, \beta ,\gamma,\delta,\kappa,\lambda$ while spacetime indices are taken from the second half, $\mu,\nu,\sigma,\rho$.}.  The metric and the connection are independent (hence the name ``Palatini").   The metric can be of any signature but we will usually take it to be of Lorentzian signature.  As a rule,  we will not write form indices, but only the internal indices, which do not mix up with the spacetime indices.  These are hidden in the forms.  It will always be clear from the context which indices are form indices and which are spacetime indices when we explicitly write the latter. 

The curvature $2$-form is defined by
\begin{equation}
R^{\beta}_{\ \alpha}(\Gamma) = d \Gamma^{\beta}_{\ \alpha} +  \Gamma^{\beta}_{\ \gamma} \wedge  \Gamma^{\gamma}_{\ \alpha} = \frac12 R^{\beta}_{\ \alpha \mu \nu} dx^\mu \wedge dx^\nu
\end{equation}
 The curvature satisfies the Bianchi identity,
\begin{eqnarray}
\nabla R^{\alpha}_{\ \beta}=  dR^{\alpha}_{\ \beta} + \Gamma^{\alpha}_{\ \gamma} \wedge R^{\gamma}_{\ \beta}- \Gamma^{\gamma}_{\ \beta} \wedge R^{\alpha}_{\ \gamma} =0    \label{Bianchi}
\end{eqnarray} 
where the operator $\nabla$ is the $GL(2n)$ covariant derivative operator with the connection ${\Gamma^\alpha}_{\beta}$.  Observe that $\nabla$ acts only on the internal indices.   The covariant derivative of the metric  $\nabla_\mu g_{\alpha \beta} \equiv  \partial_\mu g_{\alpha \beta} - {\Gamma^\gamma}_{\alpha  \mu} \, g_{\gamma \beta} - {\Gamma^\gamma}_{\beta  \mu} \, g_{\alpha \gamma}$ will in general be different from zero.

The action reads
\begin{equation}
I[g,\Gamma] = k \int  \sqrt{\vert g \vert}\, {\epsilon^{\alpha_1\alpha_2 \cdots \alpha_n}}_{\beta_1 \beta_2 \cdots \beta_n} \ R^{\beta_1}_{\ \alpha_1} \wedge R^{\beta_2}_{\ \alpha_2} \wedge \cdots \wedge R^{\beta_n}_{\ \alpha_n}\label{GB0}
\end{equation}
where $k$ is a dimensionless coupling constant and $g$ is the determinant of the metric.
We have defined 
$${\epsilon^{\alpha_1\alpha_2 \cdots \alpha_n}}_{\beta_1 \beta_2 \cdots \beta_n} \equiv g^{\alpha_1 \gamma_1} g^{\alpha_2 \gamma_2}\cdots g^{\alpha_n \gamma_n} \epsilon_{\gamma_1\gamma_2 \cdots \gamma_n \beta_1 \beta_2 \cdots \beta_n},$$
 where $\epsilon_{\gamma_1\gamma_2 \cdots \gamma_n \beta_1 \beta_2 \cdots \beta_n}$ is the $GL(2n)$  numerically invariant tensor density of weight minus one.  The combination $\sqrt{\vert g \vert}\, {\epsilon^{\alpha_1\alpha_2 \cdots \alpha_n}}_{\beta_1 \beta_2 \cdots \beta_n}$, which is the only place where the metric enters, is a $GL(2n)$ tensor acting as a multilinear form making the Lagrangian invariant under internal $GL(2n)$ linear transformations. The theory is also diffeomorphism invariant, which is  guaranteed because the whole Lagrangian is a $2n$-form. 

If the metric was fixed to be the Minkowski metric, restricting the group to the homogeneous Lorentz group, the action would be the Gauss-Bonnet topological invariant (hence the ``Gauss-Bonnet" in the terminology).  However, because the metric is varied in the action principle, the dynamics is non-trivial.

Before writing the equations of motion it is useful to introduce the following tensor density, 
\begin{eqnarray}
{\Omega^{\alpha\beta\alpha_1\alpha_2 \cdots \alpha_n}}_{\beta_1 \beta_2 \cdots \beta_n} = {\delta \over \delta g_{\alpha\beta}} \left[\sqrt{|g|}g^{\alpha_1 \gamma_1} g^{\alpha_2 \gamma_2}\cdots g^{\alpha_n \gamma_n} \epsilon_{\gamma_1\gamma_2 \cdots \gamma_n \beta_1 \beta_2 \cdots \beta_n}\right] 
\end{eqnarray} 
See Section \textbf{\ref{Sec:Lagr}} for the explicit formula at $d=4$. The equations of motion now take a simple form, 
\begin{eqnarray}
{\delta I[g,\Gamma] \over \delta g_{\alpha\beta} } &=& {\Omega^{\alpha\beta\alpha_1\alpha_2 \cdots \alpha_n}}_{\beta_1 \beta_2 \cdots \beta_n} \ R^{\beta_1}_{\ \alpha_1} \wedge R^{\beta_2}_{\ \alpha_2} \wedge \cdots \wedge R^{\beta_n}_{\ \alpha_n}=0,\\
{\delta I[g,\Gamma] \over \delta {\Gamma^{\beta}}_\alpha } &=& -n\,  {\Omega^{\gamma\kappa\, \alpha\alpha_2 \cdots \alpha_n}}_{\beta \beta_2 \cdots \beta_n} \ \nabla g_{\gamma\kappa} \wedge  R^{\beta_2}_{\ \alpha_2} \wedge R^{\beta_2}_{\ \alpha_2} \wedge \cdots R^{\beta_n}_{\ \alpha_n}=0.
\end{eqnarray}

\subsection{Gauge symmetries}
\label{Subsec:GaugeSym}

 This action possesses the following gauge symmetries,  
\begin{eqnarray}
&&\hspace{-1.2cm} \mbox{Internal GL(2n) symmetry} : \delta \Gamma^{\alpha}_{\ \beta} = \nabla \Lambda^{\alpha}_{\ \beta }, \  \  \delta g_{\alpha\beta}=\Lambda_{\alpha\beta}+\Lambda_{\beta\alpha} \\
&&\hspace{-1.2cm}\mbox{(Improved) diff invariance} : \delta \Gamma^{\alpha}_{\ \beta\mu} =- R^{\alpha}_{\ \beta\,\mu\nu}\xi^\nu,   \  \  \delta g_{\alpha\beta}= \xi^\mu \nabla_\mu g_{\alpha\beta}  \label{Eq:ImpDiff}\\
&& \hspace{-1.2cm}\mbox{Weyl symmetry}   :  \delta \Gamma^{\alpha}_{\ \beta} =0,  \ \ \ \ \ \  \ \delta g_{\alpha\beta} = 2\sigma g_{\alpha\beta},  \\
&&\hspace{-1.2cm}\mbox{Projective symmetry} : \delta \Gamma^{\alpha}_{\ \beta} = w\,\delta^{\alpha}_{\beta},  \  \ \delta g_{\alpha\beta} = 0. 
\end{eqnarray} 
where $\Lambda^{\alpha}_{\ \beta }$, $\xi^\mu$, $\sigma$ and $w$ (a $1$-form) are arbitrary spacetime functions.   These invariances are easy to check using the Bianchi identity (\ref{Bianchi}), the tracelessness of ${\epsilon^{\alpha_1\alpha_2 \cdots \alpha_n}}_{\beta_1 \beta_2 \cdots \beta_n}$ and the familiar relation $\delta R^{\mu}_{\ \nu} = \nabla \delta \Gamma^{\mu}_{\ \nu}$. Recall that the covariant derivative acts only on the internal indices.

The diffeomorphism symmetry (\ref{Eq:ImpDiff}) is presented in its ``improved" form, which is most natural in this theory. Improved diffeomorphisms appeared in \cite{JackiwTrans} as a gauge-covariant expression for coordinate transformations acting on gauge fields. Later, they became important in 2+1 Chern-Simons theories \cite{Achucarro:1987vz},\cite{Witten:1988hc},\cite{Witten:1988hf}, clarifying the relationship between the internal gauge group and diffeomorphisms. They also play an important role in higher dimensional Chern-Simons theories \cite{Banados:1995mq, Banados:1996yj}.

The main idea is to observe that a diffeomorphism can be separated into a $gl(2n)$ transformation plus an extra piece. Since the $gl(2n)$ symmetry is already accounted for, one can redefine the diffeomorphisms by keeping only the ``extra piece".  In our case, the separation for the metric and connection is,
\begin{eqnarray}
{\cal L}_{\xi} g_{\alpha \beta} &=& \xi^\nu \partial_\mu g_{\alpha \beta}  \\ 
&= &\underbrace{{}\xi^\nu \nabla_\nu g_{\alpha \beta}}_{\mbox{Improved diff}} + \underbrace{{{\Gamma^\gamma}_{\alpha  \nu}\xi^\nu \, g_{\gamma \beta}+ \Gamma^\gamma}_{\beta  \nu}\xi^\nu \, g_{\alpha \gamma}}_{\mbox{gl(2n)}}, \\
{\cal L}_{\xi} \Gamma^{\alpha}_{\ \beta\mu} &=& \xi^{\nu} \partial_\nu  \Gamma^{\alpha}_{\ \beta\mu} +    \Gamma^{\alpha}_{\ \beta\nu} \partial_\mu \xi^\nu\\
&=& \underbrace{-R^{\alpha}_{\ \beta \, \mu\nu } \xi^\nu }_{\mbox{Improved diff}} + \underbrace{\nabla_\mu(\Gamma^{\alpha}_{\ \beta\nu }\xi^\nu)}_{\mbox{gl(2n)}} .
 \end{eqnarray}  
The $gl(2n)$ transformation acts on both fields with the same parameter, $ \Gamma^{\alpha}_{\ \beta\nu }\xi^\nu$, as it should.  If the improved form of the connection transformation is used, then the metric must also transform in its improved form. 

As it can be expected, improved diffeomorphisms do not form a closed algebra. Let $\delta_X$ collectively denote a symmetry transformation with parameters $\xi^\mu, \lambda^{\alpha}_{\  \beta }, \sigma, w_\mu$ (improved diffeomorphisms, $gl(2n)$, Weyl and projective, respectively). These transformations satisfy the algebra 
\begin{eqnarray}
[\delta_{X_1},\delta_{X_2}] = \delta_{X},
\end{eqnarray} 
where the parameters at the right hand side are,
\begin{eqnarray}
\xi^\mu &=& \xi_1^{\nu}\partial_\nu \xi_2^\mu - \xi_2^{\nu}\partial_\nu \xi_1^\mu,  \\
\lambda^{\alpha}_{\ \beta} &=& -[\lambda_1,\lambda_2]^{\alpha}_{\ \beta} + R^{\alpha}_{\ \beta\mu\nu}\xi_1^\mu\xi_2^{\nu}, \\
\sigma &=& w_{1\mu} \xi_{2}^{\mu}-w_{2\mu} \xi_{1}^{\mu} + \xi_{1}^{\mu}\partial_\mu \sigma_{2}- \xi_{2}^{\mu}\partial_\mu \sigma_{1}, \\
w_\mu &=& (\partial_\mu w_{1\nu}-\partial_\nu w_{1\mu}) \xi^\nu_{2}- (\partial_\mu w_{2\nu}-\partial_\nu w_{2\mu}) \xi^\nu_{1}.
\end{eqnarray} 
We see that the background curvature acts as structure ``constants" mixing pure $gl(2n)$ transformations with improved diffs. Observe the mixing between Weyl, projective and diffeomorphisms. 

These symmetries are not independent.  If we take $\Lambda^{\alpha}_{\ \beta } = \lambda \delta^{\alpha}_{\ \beta }$, $\sigma = - \lambda$ and $w = - \nabla \lambda$, we identically get $ \delta \Gamma^{\alpha}_{\ \beta} = 0$ and  $\delta g_{\alpha\beta}=0$. Hence, we can express the dilatation part of $gl(2n)$ in terms of a Weyl transformation and a projective transformation.  Or equivalently, we can express the Weyl transformations in terms  of  $gl(2n)$ gauge transformations and projective transformations. At this stage, we do not want to commit to a particular choice and will keep the redundancy as we proceed.  This does not lead to extra work.

The above gauge symmetries are not complete either, even though they form a closed algebra.  Exhibiting the missing gauge symmetries is a harder task, which will be carried out in the Hamiltonian formulation below.

\subsection{Connection with gravity}

The internal metric $g_{\alpha\beta}$ does not define a line element in spacetime since there is no canonical way to identify the internal space of the theory with the tangent space to the manifold.  However, if a spacetime metric $g_{\mu \nu}$ (of same signature) is given, one can naturally introduce frames $\{e^\mu_\alpha\}$  that solder the internal space with the tangent space, in such a way that 
$$ g_{\alpha \beta} = e^\mu_\alpha e^\nu_\beta g_{\mu \nu} \, .$$
This is possible since the internal gauge group is $GL(2n)$.  One can impose the condition $e^\mu_\alpha = \delta^\mu_\alpha$, which makes the internal basis coincide with the basis tangent to the coordinate lines.  The action becomes then the Palatini Gauss-Bonnet term for the spacetime metric $g_{\mu \nu}$ and the spacetime connection $\Gamma^{\lambda}_{\ \mu \nu}$, which is neither metric nor symmetric.  This Palatini Gauss-Bonnet term is a natural term to be added, in the Lovelock spirit \cite{Lovelock:1971yv}, to the standard Einstein-Hilbert action in its formulation where metric and connection are treated as independent variables (one may impose the torsionless requirement if desired, see \cite{Banados:2025uhb} for the study of spherically symmetric fields in this case), just as the Chern-Simons term is a natural term to be added to the standard Yang-Mills action.  It could be the dominant one in appropriate limits.  

It is interesting to consider the pure Palatini Gauss-Bonnet action on its own.  For that reason, we shall keep the ``internal metric $GL(2n)$- connection" formulation, with the form of the gauge symmetries given above. It provides a useful approach to the gauge structure. Another reason for doing so is that generalizations to other groups $G$ can be introduced, in which one replaces the invariant metric appearing in the expression for characteristic classes by a dynamical internal metric.

Some properties of Palatini Gauss-Bonnet gravity were studied in \cite{Janssen:2019doc} and \cite{Janssen:2019uao}.   The $GL(2n)$ symmetry was not noticed though. Palatini gravity is particularly relevant in theories with high powers of the curvature tensor, for example, $f(R)$ theories. See \cite{Olmo:2011uz} for a review of this subject. The earlier paper \cite{Hehl:1994ue} remains  a valuable source of information.  The two-dimensional case was also explored in \cite{Gegenberg:1987dw}.

\subsection{Organization of the paper}
Our paper is organized as follows.  In Section {\bf \ref{Sec:2D}}, we investigate the theory in $2$ spacetime dimensions.  We show that it can be reformulated as a constrained $BF$ system.  By going to the Hamiltonian formulation, we then show  that it possesses no local degree of freedom. The presence of a new ``hidden gauge symmetry", i.e., a gauge symmetry not in the list of Subsection {\bf \ref{Subsec:GaugeSym}} is pointed out.  We then turn in Section {\bf \ref{Sec:Lagr}} to the more intricate $4$-dimensional case, for which we first analyse the equations in covariant form and construct an explicit non-trivial solution, which will be useful to test general properties of the theory. Section {\bf \ref{Sec:Ham4D}} deals with the derivation of the Hamiltonian formulation and the nature (first class versus second class) of the constraints.  In Section {\bf \ref{Sec:4DCounting}}, we show the existence of new ``hidden" gauge symmetries as in $2$ dimensions.  We also count the number of local degrees of freedom, found equal to $9$.  Finally, Section {\bf \ref{Sec:Conclusions}} deals with conclusions and prospects for future work.

\section{$2$-dimensional Palatini Gauss-Bonnet theory}
\label{Sec:2D}

We consider first the $2$-dimensional Palatini Gauss-Bonnet theory, for which the dynamics is particularly transparent and very similar to the dynamics of the Chern-Simons theory in $3$ dimensions.

\subsection{Constrained $BF$ reformulation}

In two dimensions, the action reduces to: 
\begin{equation}
I[g,\Gamma] = k \int  \sqrt{\vert g \vert}\, \epsilon^{\alpha}_{\  \beta} \, R^{\beta}_{\ \alpha}(\Gamma)  \label{GB2a}
\end{equation}
with 
\begin{equation}
\epsilon^{\alpha}_{\  \beta} = g^{\alpha \gamma} \epsilon_{\gamma \beta} \, ,
\end{equation}
We shall also set $k=1$ for notational simplicity, which is fine in the classical theory.

The Weyl rescalings
$
 \delta \Gamma^{\alpha}_{\ \beta} =0$ and $\delta g_{\alpha\beta} = 2\sigma g_{\alpha\beta} $
act algebraically.
Let us define the Weyl-invariant field
\begin{equation}
B^{\alpha}_{\  \beta} = \sqrt{\vert g \vert } \epsilon^{\alpha}_{\  \beta} = \sqrt{\vert g \vert } g^{\alpha \gamma} \epsilon_{\gamma \beta}\, .
\end{equation}
The action then takes the form of a $GL(2)$-$BF$ system \cite{Horowitz:1989ng,Birmingham:1991ty}
\begin{equation}
 \int  B^{\alpha}_{\  \beta} \, R^{\beta}_{\ \alpha}(\Gamma)  \label{GB2ab}
\end{equation}
but the $B$-field is not independent.  It follows indeed from its definition that it fulfills the conditions that its trace vanishes and that its determinant is equal to $\varepsilon$, or equivalently,
\begin{equation}
\psi \equiv B^{\alpha}_{\  \alpha} = 0 \, , \qquad \omega \equiv B^\alpha_\beta B^\beta_\alpha + 2\epsilon  = 0\, , \label{Eq:ConstraintsB}
\end{equation}
i.e., the field $B^{\alpha}_{\  \beta}$ has only two independent degrees of freedom.  Here, $\epsilon = \frac{g}{\vert g \vert}$ $=1$ (respectively $-1$) if the metric has Euclidean (respectively Lorentzian) signature.  Conversely, given a field $B^{\alpha}_{\  \beta}$ fulfilling the constraints (\ref{Eq:ConstraintsB}), one can recover 
$g_{\alpha\beta}$ up to a Weyl rescaling as
\begin{equation}
g_{\alpha\beta} = \Delta \, \epsilon_{\alpha \gamma}  B^{\gamma}_{\  \beta}
\end{equation}
where $\Delta$ is an undetermined scale factor equal to the square root $\sqrt{\vert g \vert }$.  The metric is guaranteed to be symmetric thanks to the tracelessness of $B$.

The action of two-dimensional Palatini-Gauss-Bonnet theory is thus equivalent to 
\begin{equation}
I[B,\Gamma; a,b ] = \int \Big( B^{\alpha}_{\  \beta} \, R^{\beta}_{\ \alpha}(\Gamma)  -  (a \psi + b \omega) \Big)\label{GB2b}
\end{equation}
where $a$ and $b$ are 2-form Lagrange multipliers enforcing the constraints (\ref{Eq:ConstraintsB}).  This is the standard $GL(2)$-BF action, supplemented by the constraints (\ref{Eq:ConstraintsB}).

The covariant equations of motion for 2d Palatini Gauss-Bonnet in the BF representation are,
\begin{eqnarray}
{\delta I[B,\Gamma,a,b] \over \delta B^{\alpha}_{\ \beta}} &=& R^{\beta}_{\ \alpha} - a \delta^{\beta}_{\ \alpha} - 2 b B^{\beta}_{\ \alpha}=0, \label{eqd=2B} \\
{\delta I[B,\Gamma,a,b] \over \delta \Gamma^{\alpha}_{\ \beta}} &=& \nabla B^{\beta}_{\ \alpha}=0, \label{eqd=2G}
\end{eqnarray}
plus the two constraints (\ref{Eq:ConstraintsB}).

The explicit gauge symmetries of the action (\ref{GB2b}) are $gl(2)$ transformations, diffeomorphisms, and projective transformations. The improved form for diffs acting on the B-field is,
\begin{eqnarray}
\delta_\xi B^{\alpha}_{\ \beta } = \xi^\mu \nabla_\mu B^{\alpha}_{\ \beta}.  \label{impB}
\end{eqnarray} 
Since the equations of motion determine the full curvature tensor in terms of the other fields, diffeomorphisms (in improved form) can be expressed as
\begin{eqnarray}
\delta_\xi \Gamma^{\alpha}_{\ \beta \mu } &=& -R^{\alpha}_{\ \beta\mu\nu } \xi^\nu,\\
&=& -a_{\mu\nu}\xi^\nu \delta^{\alpha}_{\ \beta } - 2b_{\mu\nu}\xi^\nu\, B^{\alpha}_{\ \beta }  .
 \end{eqnarray}
The first term is a projective transformation, already identified as a symmetry on its own. The second term is a peculiar representation of the diffeomorphism invariance  of this  theory. For the B field, note that Eq. (\ref{eqd=2G}) together with (\ref{impB}) means that diffs act trivially on $B^{\alpha}_{\ \beta}$.

We shall analyze the symmetries in detail in the Hamiltonian formalism where the classification of independent symmetries is most transparent. We anticipate some similarities with 3d Chern-Simons theory where the curvature is exactly zero making diff invariance fully contained in the internal group.

\subsection{Hamiltonian formulation}

\subsubsection{Hamiltonian action}

The action (\ref{GB2c}) is already first order and does not require additional fields to be put in Hamiltonian form.  Splitting the indices into  space and time we find the Hamiltonian form of the action,
\begin{eqnarray}
&&I[B^{\alpha}_{\  \beta},\Gamma^{\alpha}_{\ \beta\, 1}; \Gamma^{\alpha}_{\ \beta\, 0}, a,b ] \nonumber \\
&& \qquad =\int d^2 x \left(B^{\alpha}_{\  \beta }  \ (\dot \Gamma^{\beta}_{\ \alpha\, 1} - \nabla_1 \Gamma^{\beta}_{\ \alpha\, 0} ) -  a \psi - b \omega\right) \nonumber \\
&& \qquad =\int d^2 x \left(B^{\alpha}_{\  \beta }  \ \dot \Gamma^{\beta}_{\ \alpha\, 1} -  \Gamma^{\beta}_{\ \alpha\, 0} \,  \mathcal {\cal G}^{\alpha}_{\  \beta } -  a \psi - b \omega \right) \label{GB2c}
\end{eqnarray}
with 
\begin{equation}
{\cal G}^{\alpha}_{\  \beta } = - \nabla_1 B^{\alpha}_{\  \beta } \, .
\end{equation}
and we have discarded a boundary term. The phase space variables are $\Gamma^{\alpha}_{\ \beta\, 1}$ and its canonically conjugate $B^{\alpha}_{\  \beta }$,
\begin{equation}
\{ \Gamma^{\alpha}_{\ \beta\, 1}, B^{\gamma}_{\  \kappa }\} =  \delta^{\alpha}_{\  \kappa } \delta ^{\gamma}_{\  \beta }\, ,
\end{equation}
as can be seen from the ``$p \dot{q}$" kinetic term. The variables $\Gamma^{\alpha}_{\ \beta\, 0}$,  $a$ and $b$ are Lagrange multipliers enforcing respectively the constraints 
\begin{equation}
\mathcal G^{\alpha}_{\  \beta } = 0
\end{equation} 
and  (\ref{Eq:ConstraintsB}).

\subsubsection{Constraints are first class}
Since $\psi$ and $\omega$ involve only the momenta, it is clear that
\begin{equation}
\{\psi(x),     \omega(y)\} = 0 \, .
\end{equation} 

To compute the brackets involving 
$\mathcal G^{\alpha}_{\  \beta }$, we observe that these generate $GL(2)$ gauge transformations. Indeed,
\begin{eqnarray}
&&\{ \Gamma^{\alpha}_{\ \beta\, 1} , \int dy\ \mathcal G^{\gamma}_{\  \kappa } \Lambda^{\kappa}_{\  \gamma } \} = \nabla_1 \Lambda^{\beta}_{\  \alpha } \\
&&\{ B^{\alpha}_{\ \beta } , \int dy \ \mathcal G^{\gamma}_{\  \kappa } \Lambda^{\kappa}_{\  \gamma } \} =   [B, \Lambda]^{\alpha}_{\ \beta } \label{GaugeGen}
\end{eqnarray}
(matrix commutator). It follows that the $\mathcal G^{\alpha}_{\  \beta }$'s among themselves satisfies the $GL(2)$-algebra, while,
\begin{eqnarray}
\{\psi(x),     \mathcal G^{\alpha}_{\  \beta }(y) \} &=& 0,  \\
\{\omega(x),     \mathcal G^{\alpha}_{\  \beta }(y) \} &=& 0  
\end{eqnarray}
since traces are $gl(2)$ invariant.  The constraints are thus all first class.

\subsubsection{Gauge symmetries}
\label{Subsec:GaugeSym2D}

Because the constraints are first class, they generate gauge symmetries.  We already identified the gauge symmetries generated by $ \mathcal G^{\alpha}_{\  \beta }$.  It remains to understand the transformations generated by the other constraints.

It is easy to verify that
\begin{equation}
\{ \Gamma^{\alpha}_{\ \beta\, 1} , \int dy\ \varepsilon \psi \} = \varepsilon \delta^{\alpha}_{\ \beta} \, , \quad
\{ B^{\alpha}_{\ \beta } , \int dy\ \varepsilon \psi \} =   0 \, . \label{ProjGen}
\end{equation}
This is the projective symmetry expressed in phase space.  
The other symmetry reads
\begin{equation}
\{ \Gamma^{\alpha}_{\ \beta\, 1} , \int dy\ \zeta \omega \} = 2 \zeta B^{\alpha}_{\ \beta} \, , \quad
\{ B^{\alpha}_{\ \beta } , \int dy\ \zeta \omega \} =   0 \label{NewGen2}
\end{equation}
It is a new gauge symmetry, not in the list of Subsection {\bf \ref{Subsec:GaugeSym}}.

\subsection{Equations of motion and independent symmetries}
Besides the constraints, the system is subject to dynamical equations, which follow by extremizing the action with respect to the canonical variables.  These are
\begin{equation}
\nabla_0 B^{\alpha}_{\ \beta} = 0 \, , \qquad R^{\alpha}_{\ \beta 0 1} = a  \delta^{\alpha}_{\ \beta} + 2 b B^{\alpha}_{\ \beta} \label{Eq:EOM2D}
\end{equation}
where in a slight abuse of language we call $a_{01}=a$ and $b_{01}=b$. Observe that since the Hamiltonian formulation did not require the addition of extra fields, these equations coincide with the covariant ones (\ref{eqd=2B}) and (\ref{eqd=2G}).  

Among the gauge symmetries  of Subsection {\bf \ref{Subsec:GaugeSym}}, we have already identified the Hamiltonian expressions of the $GL(2)$ internal gauge transformations and of the projective transformations. This is also trivially true for Weyl invariance, which has indeed no action in the BF-formulation since the variables are Weyl invariant.

The question is then: what is the canonical expression of the spacetime diffeomorphisms?  The spatial ones act also trivially in phase space since $R^{\alpha}_{\ \beta 1 1} = 0$ and $\nabla_1 B^{\alpha}_{\ \beta} = 0$.  As to the time diffeomorphisms, Eq. (\ref{Eq:EOM2D}) shows that they
 can be expressed in terms of the projective gauge transformations and of the new gauge symmetry.

Hence, we have not only accounted for all the gauge symmetries of subsection {\bf \ref{Subsec:GaugeSym}}, but we have also found a new one, namely (\ref{NewGen2}).  One might try to express (\ref{NewGen2}) as a combination of the gauge symmetries of subsection {\bf \ref{Subsec:GaugeSym}}, but this necessitates taking $\xi^0 = \zeta/b$, which cannot be done if the Lagrange multiplier $b$ vanishes. Without extra consideration, there is no reason to impose $b \not=0$. Therefore, it seems more appropriate to regard (\ref{NewGen2}) as a genuine new gauge symmetry not in the list of {\bf \ref{Subsec:GaugeSym}} and consider the time diffeomorphisms as dependent on it.  The fact that spacetime diffeomorphisms are not independent gauge symmetries is of course not unexpected; this property also holds for the standard $BF$ system as well as for Chern-Simons theory in $3$ spacetime dimensions.

\subsection{Counting the number of degrees of freedom}
The number of first class constraints is $4 + 1 + 1$ but they cannot be independent because the number of independent conjugate pairs is $4$.  

In fact there are exactly $4$ independent first class constraints (and not less). Directly from the definitions one finds two relations among the constraints, 
\begin{eqnarray}
\nabla \psi + {\cal G}^{\alpha}_{\ \alpha}&=&0, \\
 \nabla \omega + 2 B^{\alpha}_{\ \beta}\,{\cal G}^{\beta}_{\  \alpha}&=&0
\end{eqnarray} 
The counting of degrees of freedom, number of canonical fields - 2$\times$(independent first class constraints) is then, 
\begin{eqnarray}
2 \times 4 - 2 \times 4 =0.
\end{eqnarray} 

One can also show that,  by using the gauge transformations and the constraints, one can bring the canonical variables to some definite set of fixed values.

This can be seen as follows.  The matrix $B$ being traceless and of determinant $1$ (because of the constraints)\footnote{We consider for definiteness the case of Euclidean signature ($\epsilon = 1$).  If the case of Lorentzian signature ($\epsilon = -1$), the eigenvalues are $\pm 1$ and $\sigma_3$ plays then the role played by $i\sigma_2$ in the text.} has pure imaginary eigenvalues $\lambda_1$, $\lambda_2$ equal to $\pm i$.  By a real transformation one cannot diagonalize it, but one can bring it to the form $i \sigma_2$.  Once this is done, $\Gamma_1$ (the matrix with matrix elements $ \Gamma^{\alpha}_{\ \beta\, 1}$) must commute with $\sigma_2$, i.e., must take the form $\Gamma_1 = \alpha I + \beta (i \sigma_2)$.  Similarly, the residual gauge transformations  must also commute with $\sigma_2$ and read $\Lambda = \rho I + \sigma (i \sigma_2)$. This implies that they have no action on $\Gamma_1$.  

However, the projective gauge transformations and the new gauge transformations do act on $\Gamma_1$.  The projective transformations can be used to set $\alpha = 0$, while the new  transformations can be used to set $\beta = 0$.

Hence by using all the gauge freedom and the constraints, one can bring the canonical variables to the form
\begin{equation}
\Gamma_{ 1} = 0 \, , \qquad B = i \sigma_2 \, .
\end{equation}
The theory has therefore zero degree of freedom.

\subsection{The role of the extra constraints}
It might seem that the extra constraints (\ref{Eq:ConstraintsB}) (with respect to those of the standard $BF$-system) make the theory completely trivial.      Consider for instance the case where the manifold is $\mathbb{R} \times S^1$.  The standard $BF$ model contains then a finite number of global degrees of freedom, which can be viewed as the eigenvalues of the $\Gamma$-holonomy around the circle and the eigenvalues of the $B$-field in a frame where the $B$-field is constant \cite{Horowitz:1989ng,Birmingham:1991ty}.  These define orbits of the $GL(2)$ action. Now, we just saw that these eigenvalues are completely fixed by the constraints  (\ref{Eq:ConstraintsB}) while the eigenvalues of the connection can be changed and set to zero by the new gauge symmetries. So this theory is restricted to a single orbit and has neither local nor global degrees of freedom.

It turns out that this conclusion does not hold on manifolds with boundaries.  This will be explored in a forthcoming work \cite{Banados:2026}.

\section{4-dimensional Palatini Gauss-Bonnet theory - Covariant formulation}
\label{Sec:Lagr}

\subsection{Action and equations of motion}

In $4$ spacetime dimensions,  the Palatini Gauss-Bonnet action reads
\begin{equation}
I[g,\Gamma] = \int  \sqrt{g}\, \epsilon^{\alpha\gamma}_{\ \ \ \beta\delta} \ R^{\beta}_{\ \alpha}(\Gamma) \wedge R^{\delta}_{\ \gamma}(\Gamma) \label{GB}
\end{equation}

The equations of motion that follow from the Lagrangian are, 
\begin{eqnarray}
\Omega^{\kappa\lambda\, \alpha \gamma}_{\ \ \ \ \ \beta \delta} \, R^{\beta}_{\ \alpha} \wedge R^{\delta}_{\ \gamma} =0 \label{covg} \\
\Omega^{\kappa\lambda\, \alpha \gamma}_{\ \ \ \ \ \beta \delta} \, R^{\beta}_{\ \alpha} \wedge \nabla g_{\kappa\lambda}  =0 \label{covG}
\end{eqnarray} 
where 
\begin{eqnarray}
\Omega^{\kappa\lambda\, \alpha \gamma}_{\ \ \ \ \ \beta \delta} &\equiv & {\delta  \over  \delta g_{\kappa\lambda} } \Big( \sqrt{g}\, \epsilon^{\alpha\gamma}_{\ \ \ \beta\delta}\Big)  \\
&=& {1 \over 2}\sqrt{g} \left(   g^{\kappa\lambda}\epsilon^{\alpha\gamma}_{\ \ \ \beta\delta} - g^{\alpha\sigma}\epsilon^{\rho\gamma}_{\ \ \ \beta\delta} -g^{\alpha\rho}\epsilon^{\sigma\gamma}_{\ \ \ \beta\delta}-g^{\gamma\rho}\epsilon^{\alpha\sigma}_{\ \ \ \beta\delta}-g^{\gamma\sigma}\epsilon^{\alpha\rho}_{\ \ \ \beta\delta} \right). \nonumber
\end{eqnarray} 
Here we have used 
\begin{equation}
{\delta g_{\alpha\beta}  \over \delta g_{\gamma\delta}} = {1 \over 2} \Big(\delta^\gamma_\alpha\delta^\delta_\beta +\delta^\delta_\alpha\delta^\gamma_\beta\Big).
\end{equation}

The tensor $\Omega^{\kappa\lambda\, \alpha \gamma}_{\ \ \ \ \ \beta \delta}$ inherits several properties from the Levi-Cevita density like vanishing traces, 
\begin{equation}
\Omega^{\kappa\lambda\, \alpha \gamma}_{\ \ \ \ \ \alpha \delta} = \Omega^{\kappa\lambda\, \alpha \gamma}_{\ \ \ \ \ \beta \gamma}= 0 \, ,
\end{equation}
antisymmetry under $\alpha\leftrightarrow \gamma$ and $\beta \leftrightarrow \delta$ and symmetry under simultaneous interchange of those indices. Furthermore, it satisfies,
\begin{eqnarray}
g_{\kappa\lambda}\Omega^{\kappa\lambda\, \alpha \gamma}_{\ \ \ \ \ \beta \delta}&=&0, \\
\Omega^{\sigma \  \alpha\gamma}_{\ \rho \ \ \   \beta\delta} +\Omega^{\alpha \  \sigma\gamma}_{\ \beta \ \ \   \rho\delta}+\Omega^{\gamma \  \alpha\sigma}_{\ \delta \ \ \   \beta\rho} &=& 0.
 \label{omegaid}
\end{eqnarray}
The vanishing of the first trace descends from Weyl invariance; the second identity follows from the fact that  $g^{\lambda[\kappa}\epsilon^{\alpha \gamma  \beta \delta]}= 0$ (there is no totally antisymmetric tensor with $5$ indices in $4$ dimensions).  Indeed one finds by direct computation (raising all indices with the metric) that
\begin{eqnarray}
&& \hspace{-1cm} \Omega^{\lambda \kappa  \alpha\gamma   \beta\delta} +\Omega^{\alpha \beta \lambda\gamma  \kappa\delta}+\Omega^{\gamma \delta  \alpha\lambda  \beta\kappa}  \nonumber \\
&&=  g^{\kappa[\lambda}\epsilon^{\alpha \gamma  \beta \delta]} + g^{\beta[\alpha}\epsilon^{\lambda \gamma  \kappa \delta]}  + g^{\delta[\gamma}\epsilon^{\alpha \lambda  \beta \kappa]} = 0 \, .
\end{eqnarray}

Note that by definition of the tensor density $\Omega$, it follows,
\begin{eqnarray}
&& \delta (\sqrt{g}\, \epsilon^{\alpha\gamma}_{\ \ \ \beta\delta}) = \delta g_{ \kappa\lambda } \,  \Omega^{\kappa\lambda\, \alpha \gamma}_{\ \ \ \ \ \beta \delta}\\
&& \nabla (\sqrt{g}\, \epsilon^{\alpha\gamma}_{\ \ \ \beta\delta})  = \nabla g_{ \kappa\lambda } \,  \Omega^{\kappa\lambda\, \alpha \gamma}_{\ \ \ \ \ \beta \delta}
\end{eqnarray}

A nice check of the  equations of motion is as follows: For the Levi-Civita connection, the curvature becomes the Riemann tensor and $\nabla g_{\kappa\lambda}=0$. On that sector of field space, both (\ref{covg}) and (\ref{covG}) must be trivial since the Lagrangian is a total derivative. Eq.  (\ref{covG}) is indeed identically satisfied. Eq. (\ref{covg}) is far less obvious but still identically satisfied thanks to a so called ``topological identity" \cite{lovelock1970dimensionally,Schwimmer:2000cu}.

The equation for the connection (\ref{covG}) can also be written as 
\begin{eqnarray}
\nabla ^*R^{\alpha}_{\ \beta}=0 \label{self}
\end{eqnarray} 
where
\begin{eqnarray}
^*R^{\alpha}_{\ \beta} = {1 \over 2} \varepsilon^{\alpha\gamma}_{\ \ \ \beta\delta}R^{\delta}_{\ \gamma} 
\end{eqnarray}  
is the dual (in internal space) of $R^{\alpha}_{\ \beta} $.  One can check that the covariant derivative of this equation gives exactly the metric equation (\ref{covg}). Thus, the metric equation is not independent. Once the connection equation has been solved the whole system has been solved.

The form (\ref{self}) has some applications.  First, Euclidean selfdual fields   $^*R^{\alpha}_{\ \beta}=R^{\alpha}_{\ \beta} $, for which (\ref{self}) is mapped on the Bianchi identity (\ref{Bianchi}),  are automatically solutions. This is the same mechanism for building selfdual solutions to Yang-Mills and gravitational theories. Second, one may wonder if a cosmological constant $\Lambda\int \sqrt{g}$ could be added to this theory. Such a term would break $GL(4)$ to $SL(4)$, but apart from that it is an acceptable term. If $\Lambda$ is added, then a new term appears at the right hand side of (\ref{covg}). However, it turns out that the integrability condition on (\ref{self}) implies $\Lambda=0$. 

\subsection{A particular solution}
\label{Sec:Part}

As we shall discuss below in great detail, the Palatini Gauss-Bonnet theory has the feature that the rank of the matrix of the Poisson brackets of the constraints is not constant over the constraint surface. The constraint surface can then be written as the union of regions on each of which the rank is constant.  The highest possible value of the rank compatible with the constraints will be called the maximum rank, and corresponds to the generic case since lower rank regions are defined by additional equations stating that there are additional zero eigenvalues,  and are therefore submanifolds. 

We shall call ``good background"  a solution of the equations of motion that precisely lies in the maximum rank region.
Finding a good background in a theory with variable rank can be quite challenging. Indeed, if one simplifies the equations too much by imposing symmetries, the danger is to hit non generic, lower rank regions.  One must keep sufficient complexity. For Palatini Gauss-Bonnet theory, multilinearity of the equations of motion  plays an important role to make this task somehow simpler and families of solutions can be found.

For example, one solution of the equations of motion corresponding to a good background is the Minkowski metric
\begin{equation}
g = (g_{\alpha \beta}) = \begin{pmatrix} -1 & 0 & 0 & 0 \\
0 & 1 & 0 & 0 \\
0 & 0 & 1 & 0 \\
0 & 0 & 0 & 1 
\end{pmatrix}
\end{equation}
with the constant connection $1$-form
\begin{equation}
\Gamma = \begin{pmatrix} 2dz & dy & dx & dy \\ 
dy & dx+dz & 2 dy &-3 dx \\
dx & 2dy &3 dx-dz &dy \\
dy &-3 dx &dy &-4 dx -2 dz
\end{pmatrix}
\end{equation}
One finds
\begin{equation}
\nabla g = \begin{pmatrix} 0 & 0 &0 & 0 \\ 
0 & 2dx+2 dz & 4 dy &-6 dx \\
0 & 4dy &6 dx-2 dz &2 dy \\
0 &-6 dx &2dy &-8 dx -4 dz
\end{pmatrix}
\end{equation}
and (wedge omitted)
\begin{equation}
R = \begin{pmatrix} 0 & - dy  dz + 4 dx  d y &3 dz  d x & -4 dy  dz + 8 dx  dy \\ 
dy  dz - 4 dx  dy  & 0 & -4 dy  dz - 8 dx  dy &- 9 dz  dx \\
- 3 dz dx & 4dy dz + 8 dx dy &0 &-dydz + 14 dx dy \\
4 dy dz - 8 dx dy &9 dz dx &dydz - 14 dx dy &0
\end{pmatrix} 
\end{equation}
The connection is chosen to be symmetric in order that the covariant derivatives of the spatial metric are ``maximally non-zero''.  The curvature is antisymmetric because it is given by commutators of symmetric matrices. 

An algebraically direct but somewhat involved computation shows that these constant metric and connection solve the equations of motion. This solution will be useful below because its gives indeed, as we shall see, a non trivial example of a solution with ``maximum rank''. See \cite{Banados:2025uhb} for other solutions and the study of this theory when the connection is symmetric. 

\section{4-dimensional Palatini Gauss-Bonnet theory - 3+1 decomposition and Hamiltonian formulation}
\label{Sec:Ham4D}
 
\subsection{3+1 decomposition}
In order to derive the Hamiltonian formulation, we need to perform the 3+1  decomposition of the spacetime objects.  This affects only the connection and not the metric in internal space, which corresponds to a collection of scalars from the spacetime point of view.  So we write
\begin{eqnarray}
\Gamma^{\alpha}_{\  \beta} &=& \Gamma^{\alpha}_{\  \beta \mu} dx^\mu \\
&=& \Gamma^{\alpha}_{\  \beta 0} dt + \Gamma^{\alpha}_{\  \beta i} dx^i \label{3p1}
\end{eqnarray}

The space-space components of the curvature are,
\begin{eqnarray}
R^{\alpha}_{\ \beta\, ij} = \partial_i \Gamma^{\alpha}_{\  \beta i}-\partial_j \Gamma^{\alpha}_{\  \beta j} +  \Gamma^{\alpha}_{\  \gamma i} \Gamma^{\gamma}_{\  \beta j}-\Gamma^{\alpha}_{\  \gamma j} \Gamma^{\gamma}_{\  \beta i} 
\end{eqnarray} 
The time-space components contain the only time derivative in this theory, 
\begin{eqnarray}
R^{\alpha}_{\ \beta\, 0i} &=& \partial_0 \Gamma^{\alpha}_{\  \beta i}-\partial_i \Gamma^{\alpha}_{\  \beta 0} +  \Gamma^{\alpha}_{\  \gamma 0} \Gamma^{\gamma}_{\  \beta i}-\Gamma^{\alpha}_{\  \gamma j} \Gamma^{\gamma}_{\  \beta 0} \\
&=&   {d\Gamma^{\alpha}_{\  \beta i} \over dt} - \nabla_{i} \Gamma^{\alpha}_{\  \beta 0}
\end{eqnarray} 
 
The Lagrangian becomes
\begin{eqnarray}
{\cal L} &=& \sqrt{g}\epsilon^{\alpha \gamma}_{\ \ \ \beta \delta} \ R^{\beta}_{\ \alpha\,0i} R^{\delta}_{\ \gamma\,jk}   \epsilon^{ijk} \\
&=& \sqrt{g}\ \epsilon^{\alpha \gamma}_{\ \ \ \beta \delta}  \ (\dot \Gamma^{\alpha}_{\ \beta\, i} - \nabla_i \Gamma^{\alpha}_{\ \beta\, 0} ) R^{\delta}_{\ \gamma\,jk}   \epsilon^{ijk} 
\end{eqnarray}

The Lagrangian equations of motion written in a 3+1 decomposition will be useful below.  Applying (\ref{3p1}) to (\ref{covg}) and (\ref{covG}) we find three equations:
\begin{eqnarray}
\Omega^{\kappa\lambda\, \alpha \gamma}_{\ \ \ \ \ \beta \delta} \epsilon^{ijk} R^{\beta}_{\ \alpha ij}R^{\delta}_{\ \gamma k0}&=& 0, \label{3p11} \\
 \Omega^{\kappa\lambda\, \alpha \gamma}_{\ \ \ \ \ \beta \delta} \epsilon^{ijk} (R^{\beta}_{\ \alpha ij}\nabla_0 g_{\kappa\lambda} -2 \nabla_i g_{\kappa\lambda} R^{\beta}_{\ \alpha j0 } )&=& 0, \label{3p12}\\
\Omega^{\kappa\lambda\, \alpha \gamma}_{\ \ \ \ \ \beta \delta} \epsilon^{ijk} R^{\beta}_{\ \alpha ij}\nabla_{k}g_{\kappa\lambda}  &=& 0. \label{3p13}
\end{eqnarray} 
Eq. (\ref{3p11}) is identical to (\ref{covg}). Equations (\ref{3p12}) and (\ref{3p13}) represent (\ref{covG}) in a 3+1 form. 

One can set up an initial value problem in which one would give on a ``Cauchy'' hypersurface $t=0$ the spatial components $\Gamma^{\alpha}_{\ \beta\, i}$ of the connection and the metric $g_{\alpha \beta}$ subject to the constraints (\ref{3p13}).  If one were to know the $GL(4)$-invariant  time derivatives of these fields, i.e.,  $R^{\delta}_{\ \gamma k0}$ and $\nabla_0 g_{\kappa\lambda}$, one could  determine their evolution off the initial hypersurface up to $GL(4)$ gauge transformations.  These time derivatives are to be extracted from the ``dynamical equations''  (\ref{3p11}) and (\ref{3p12}).  

One can easily check that whatever be $R^{\delta}_{\ \gamma k0}$ and $\nabla_0 g_{\kappa\lambda}$ fulfilling  (\ref{3p11}) and (\ref{3p12}), the constraints (\ref{3p13}) are preserved by the time evolution, i.e., 
$$
\partial_0 \left(\Omega^{\kappa\lambda\, \alpha \gamma}_{\ \ \ \ \ \beta \delta} \epsilon^{ijk} R^{\beta}_{\ \alpha ij}\nabla_{k}g_{\kappa\lambda}\right) \simeq 0 \, ,
$$
if one takes into account (\ref{3p11}), (\ref{3p12}) and (\ref{3p13}) {\em undifferentiated with respect to time} (this is the meaning of the symbol $\simeq$).  Indeed, this equation is equivalent to $$\nabla_0 \left(\Omega^{\kappa\lambda\, \alpha \gamma}_{\ \ \ \ \ \beta \delta} \epsilon^{ijk} R^{\beta}_{\ \alpha ij}\nabla_{k}g_{\kappa\lambda}\right)\simeq 0$$ if we make use of (\ref{3p13}).   But this is the same as
$$ 
 \nabla_0\left(\frac{\delta \mathcal L}{\delta {\Gamma^\alpha}_{\beta 0}}\right)\simeq 0
  $$ 
which indeed is zero if we recall the Noether identities following from the gauge invariance,
$$
  \nabla_\mu\left(\frac{\delta \mathcal L}{\delta {\Gamma^\delta}_{\gamma \mu}}\right)-  
 \frac{\delta \mathcal L}{\delta g_{\alpha \beta} } (g_{\alpha \delta} \delta^\gamma_\beta + g_{\beta \delta} \delta^\gamma_\alpha)= 0 \, ,
 $$ 
 which can be rewritten as
$$
  \nabla_0\left(\frac{\delta \mathcal L}{\delta {\Gamma^\delta}_{\gamma 0}}\right) = -  \nabla_k\left(\frac{\delta \mathcal L}{\delta {\Gamma^\delta}_{\gamma k}}\right) +
 \frac{\delta \mathcal L}{\delta g_{\alpha \beta} } (g_{\alpha \delta} \delta^\gamma_\beta + g_{\beta \delta} \delta^\gamma_\alpha)= 0 \, ,$$ 
 or
$$
 `` \nabla_0 (\ref{3p13}) = - \boldsymbol{\nabla} (\ref{3p12}) +(\ref{3p11})" \, .
$$
Therefore, there are no further constraints. 

The question, then, is to understand how the dynamical equations  (\ref{3p11}) and (\ref{3p12}) determine $R^{\delta}_{\ \gamma k0}$ and $\nabla_0 g_{\kappa\lambda}$.   
This issue is best analyzed in the Hamiltonian formalism, where the dynamical equations are put in an equivalent normal form, with time derivatives on the left-hand side and everything else, including the Lagrange multipliers, on the right-hand side.

\subsection{Hamiltonian action}
The time component of the connection enters the Lagrangian linearly and plays the role of a Lagrange multiplier.  For that reason, we define momenta $\phi^{i\beta}_{\ \ \alpha}$ and $\pi^{\alpha \beta}$ only for the space components $\Gamma^{\alpha}_{\ \beta i }$ and the metric $g_{\alpha\beta}$. This leads to two primary constraints,
\begin{eqnarray}
\pi^{\alpha \beta} &=&0  \label{pi} \\
\phi^{i\beta}_{\ \ \alpha} &=& p^{i\beta}_{\ \alpha } - \sqrt{g}\epsilon^{\beta \gamma}_{\ \ \ \alpha  \delta} \   R^{\delta}_{\ \gamma\,jk}\epsilon^{ijk}   \label{phi}
\end{eqnarray} 

The Hamiltonian action reads, 
\begin{eqnarray}
&&\hspace{-.7cm} I[\Gamma^{\beta}_{\ \alpha\, i }, p^{i\alpha}_{\ \, \beta}, g_{\alpha\beta}, \pi^{\alpha\beta}; \Gamma^{\beta}_{\ \alpha 0 }, X^{\alpha}_{\ \beta\, i}, Y_{\alpha\beta}] \nonumber \\
&&= \int p^{i\alpha}_{\ \, \beta} \dot \Gamma^{\beta}_{\ \alpha\, i }+ \pi^{\alpha\beta}\dot g_{\alpha\beta}  - (\Gamma^{\beta}_{\ \alpha 0 }\, {\cal G}^{\alpha}_{\ \beta}  + X^{\alpha}_{\ \beta\, i} \phi^{i \beta}_{\ \alpha}  + Y_{\alpha\beta} \pi^{\alpha\beta})
\end{eqnarray} 
with
\begin{eqnarray}
{\cal G}^{\alpha}_{\ \beta} = -\nabla_{i} p^{i\alpha}_{\ \ \beta } + 2 \pi^{\alpha\gamma} g_{\gamma\beta}  \, . \label{Gauss} 
\end{eqnarray} 
We have included in the Gauss constraint the weakly vanishing term $ 2 \pi^{\alpha\sigma} g_{\sigma\beta}$ such that it generates the complete internal $GL(4)$ gauge symmetry on all the variables, including on the metric.
Note that while the Lagrange multiplier $\Gamma^{\alpha}_{\ \beta0 }$ was already present in the Lagrangian action, the variables $X^{\alpha}_{\ \beta i}$ and $Y_{\alpha\beta}$ are new Lagrange multipliers introduced to enforce the two primary constraints. 

Varying the action with respect to the Lagrange multipliers $X^{\alpha}_{\ \beta\, i}$ and $ Y_{\alpha\beta}$ yields the constraints $\pi^{\alpha \beta} = 0$ and $\phi^{i \beta}_{\ \alpha} = 0$, which can clearly be solved for the momenta.  Varying the action with respect to the momenta yields expressions of the Lagrange multipliers in terms of the time derivatives of the Lagrangian variables, 
\begin{eqnarray}
&& \dot{g}_{\alpha \beta} = \Gamma^{\gamma}_{\ \alpha 0 } \, g_{\gamma \beta} + \Gamma^{\gamma}_{\ \beta 0 } \, g_{\alpha \gamma } + Y_{\alpha \beta}= 0 \quad \Leftrightarrow \quad Y_{\alpha \beta} = \nabla_0 g_{\alpha \beta} \label{Eq:Y}\\
&&\dot \Gamma^{\beta}_{\ \alpha\, i } = \nabla_i \Gamma^{\beta}_{\ \alpha 0 } + X^{\alpha}_{\ \beta\, i} = 0 \quad \Leftrightarrow \quad X^{\alpha}_{\ \beta\, i} =  R^{\alpha}_{\ \beta 0i  } \label{Eq:X}
\end{eqnarray}

Thus the pairs $p^{i\alpha}_{\ \beta},X^{\alpha}_{\ \beta i}$ and $\pi^{\alpha\beta},Y_{\gamma\delta} $ form auxiliary fields that can be solved for algebraically using their own equations of motion.  By eliminating these fields from the Hamiltonian action, the original covariant action is recovered.   

The remaining equations are the equations obtained by extremizing the Hamiltonian action with respect to the Lagrange multiplier $\Gamma^{\alpha}_{\ \beta0 }$, the metric and the spatial components $\Gamma^{\beta}_{\ \alpha\, i }$ of the connection.

Extremizing the Hamiltonian action with respect to the Lagrange multiplier $\Gamma^{\alpha}_{\ \beta0 }$ gives the Gauss constraint:
\begin{equation}
{\cal G}^{\alpha}_{\ \beta} = 0 \, ,
\end{equation}
which is clearly equivalent to (\ref{3p13}) taking $\pi^{\alpha \beta} = 0$ and the definition of the momenta $p^{i\alpha}_{\ \beta}$ into account.  Extremizing the action with respect to the metric reproduces  (\ref{3p11}) whereas extremizing the action with respect to the spatial components $\Gamma^{\beta}_{\ \alpha\, i }$ of the connection gives (\ref{3p12})
(taking the primary constraints and the expressions of the Lagrange multipliers into account).

\subsection{Contraint algebra, symmetries, reducibility}
The above constraints satisfy the Poisson bracket relations, 
\begin{eqnarray}
~[{\cal G}^{\alpha}_{\ \beta},{\cal G}^{\gamma}_{\ \delta}] &=& \delta^{\gamma}_{\ \beta} {\cal G}^{\alpha}_{\ \delta}-\delta^{\alpha}_{\ \delta} {\cal G}^{\gamma}_{\ \beta}  \label{Eq:PBC1}\\
~ [{\cal G}^{\alpha}_{\ \ \beta},\phi^{i \gamma}_{\ \ \delta}] &=& \delta^{\gamma}_{\beta }\phi^{i \alpha}_{\ \ \delta }-\delta^{\alpha}_{\delta }\phi^{i \gamma}_{\ \ \beta }  \label{Eq:PBC2}\\
~[{\cal G}^{\alpha}_{\ \beta},\pi^{\gamma\delta}] &=& \pi^{\alpha\delta}\delta^{\gamma}_{\beta}+\pi^{\gamma\alpha}\delta^{\delta}_{\beta} 
\label{Gpi} \\
~ [\phi^{i\alpha}_{\ \ \beta },\phi^{i\gamma}_{\ \ \delta }] &=& \Omega^{\kappa\lambda\alpha\gamma}_{\ \ \ \ \ \beta\delta} \epsilon^{ijk}\nabla_k g_{\kappa\lambda} \label{fp} \\
~  [\phi^{i\alpha}_{\ \ \beta },\pi^{\kappa\lambda }] &=& \Omega^{\kappa\lambda\alpha\gamma}_{\ \ \ \ \ \beta\delta}\epsilon^{ijk} R^{\delta}_{\ \gamma\,  jk}  \label{Eq:PBC5}\\    
~[\pi^{\alpha\beta},\pi^{\gamma\delta}]&=&0     \label{Eq:PBC6}
\end{eqnarray} 
Among these constraints, some are first class and generate the gauge symmetries while some are second class.  A difficulty is that the rank of the matrix of the Poisson brackets depends on the location on the constraint surface.  For instance, if the connection vanishes and the metric is constant - which is evidently a solution of the equations of motion - , the brackets of the constraints all weakly vanish.  On the other hand for less trivial solutions, some of the brackets do not vanish, even weakly, implying the presence of second class constraints.

The situation is similar to that encountered in \cite{Banados:1995mq,Banados:1996yj} for higher-dimensional pure Chern-Simons theories.  Just as in those references, we shall focus here our attention on the generic situation where the rank of the matrix of the Poisson brackets takes the maximum possible value compatible with the constraints.  We recall that ranks smaller than the maximum value that can be attained on the constraint surface are non generic because they are characterized by equations and hence define submanifolds of the constraint surface.

We begin with the discussion of the generators of the gauge symmetries listed above in the covariant description, which define first-class constraints no matter what the rank of the Poissson brackets of the constraints is.  Some of them can be identified easily, the others  are ``hidden" in the constraints, $\phi^{i\alpha}_{\ \ \beta}$ and $\pi^{\alpha\beta}$.

\begin{enumerate}
\item  Internal $GL(4)$ transformations: these are generated by ${\cal G}^{\alpha}_{\ \beta} $.  These constraints are clearly first class as expressed by (\ref{Eq:PBC1})-(\ref{Gpi}).  Equation (\ref{Eq:PBC1}) reproduces the $gl(4)$-algebra, while (\ref{Eq:PBC2})-(\ref{Gpi}) indicate how $\phi^{i\alpha}_{\ \ \beta}$ and $\pi^{\alpha\beta}$ (homogeneously) transform under $gl(4)$ transformations.
\item Weyl symmetry: it is generated by
\begin{equation}
W := g_{\alpha\beta}\pi^{\alpha\beta} \approx 0. 
\end{equation}
which is also first class. Indeed, it commutes weakly with ${\cal G}^{\alpha}_{\ \beta}$ and with itself. The only non-trivial relation is (\ref{fp}). However, recall from (\ref{omegaid}) that $g_{\gamma\delta}\Omega^{\gamma\delta\, \alpha\sigma}_{\ \ \ \ \ \ \beta\rho}=0$, so the trace of $\pi^{\alpha\beta} $ is indeed first class. It is worth stressing again that this trace property of $\Omega$ descends from the Weyl invariance of the action. 
\item Projective symmetry: the spatial part of the projective symmetry, parametrized by $w_i$, is generated by the trace
\begin{equation}
P^i :=\phi^{i\alpha}_{\ \alpha}  \approx 0
\end{equation}
which can be seen to be first class due the tracelessness of $\Omega$.  The temporal part parametrized by $w_0$ acts only on the Lagrange multipliers $\Gamma^{\alpha}_{\ \beta0 }$ and has therefore no constraint associated with it. It arises because the constraints are not independent.
\item Spatial diffeomorphisms: these are generated by the combination
\begin{equation}
H_i := \phi^{j\alpha}_{\ \ \beta}\, R^{\beta}_{\ \rho ji} + \pi^{\alpha\beta}\nabla_i g_{\alpha\beta}  \approx 0
\end{equation}
This generator satisfies the diffeomorphism algebra (of the improved diffeomorphisms), and it can easily be verified to commute weakly with all the other constraints. 

The generator of time diffeomorphisms will be discussed below.
\end{enumerate}      

That the constraints are not all independent (``reducible'') is easy to verify since one has identically 
\begin{equation}
\nabla_i\phi^{i\alpha}_{\ \alpha } + {\cal G}^{\alpha}_{\ \alpha} -2 \pi^{\alpha}_{\ \alpha}=0.
\label{relc}
\end{equation}
This relation just reflects the reducibility found for the symmetries in their covariant form. It is easy to see that a Weyl transformation can be generated as a projective plus a gauge transformation. 

Since the constraints are reducible, one can transform the Lagrange multipliers without changing the action,
\begin{eqnarray}
&&\Gamma^{\beta}_{\ \alpha 0 } \rightarrow \Gamma^{\beta}_{\ \alpha 0 } + w_0 \delta^\beta_\alpha \label{Eq:HPro}\\
&&X^{\alpha}_{\ \beta\, i} \rightarrow X^{\alpha}_{\ \beta\, i} - \nabla_i w_0 \delta^\beta_\alpha\\
&&Y_{\alpha\beta} \rightarrow Y_{\alpha\beta} + w_0 g_{\alpha \beta}
\end{eqnarray}
As announced, this accounts through (\ref{Eq:HPro}) for the temporal part of the projective symmetry.  The transformation of the Lagrange multipliers $X^{\alpha}_{\ \beta\, i}$ and $Y_{\alpha\beta}$ can be understood in terms of their relation with the time derivatives of the fields.

One can eliminate the redundancy of the gauge symmetries either by restricting the gauge group $GL(4)$ to $SL(4)$ and keeping the Weyl symmetry as an independent symmetry, or by keeping the full $GL(4)$ and dropping the Weyl symmetry as an independent symmetry.  Either choice fixes the redundancy in the Lagrange multipliers.  There is, however, no particular advantage in eliminating the redundancy, which we shall therefore keep along in the subsequent discussion.

One may ask the question as to whether there are other, more subtle, reducibility identities. i.e., are there other combinations of the constraints besides (\ref{relc}) that vanish identically?  There exist 16 combinations of the constraints (and only 16)  that do not contain the momenta,
namely,
\begin{equation}
\nabla_i\phi^{i\alpha}_{\ \beta } + {\cal G}^{\alpha}_{\ \beta} -2 \pi^{\alpha}_{\ \beta}= \Omega^{\kappa\lambda\, \gamma \alpha}_{\ \ \ \ \ \delta \beta } \epsilon^{ijk} R^{\delta}_{\ \gamma ij}\nabla_{k}g_{\kappa\lambda},
\end{equation}
of which the trace gives (\ref{relc}).  These equations are actually just (\ref{3p13}).  So the question is whether the equations $\Omega^{\kappa\lambda\, \gamma \alpha}_{\ \ \ \ \ \delta \beta } \epsilon^{ijk} R^{\delta}_{\ \gamma ij}\nabla_{k}g_{\kappa\lambda} = 0$ are 15 independent equations.  If this is not the case, there would be more reducibility relations among the constraints.

We have checked that the equations $\Omega^{\kappa\lambda\, \gamma \alpha}_{\ \ \ \ \ \delta \beta } \epsilon^{ijk} R^{\delta}_{\ \gamma ij}\nabla_{k}g_{\kappa\lambda} = 0$ are generically $15$ independent equations (at each spatial point).  One way to proceed is to consider the particular solution of Section {\bf \ref{Sec:Part}} and perturb it.  At any given point, the matrix of the linear system that constrains $\delta R^{\delta}_{\ \gamma ij}$ and $\delta \nabla_{k}g_{\kappa\lambda}$ (which can otherwise be viewed as independent variables at that point) is of maximum rank $15$.  Because this is the maximum possible rank, the result is generically true, in the sense that it is stable under small deformations of the background.

Given that there is no further reducibility identity on the constraints, we can conclude that there are 
$$ 16+3 \times 16 + 10 - 1 = 73
$$
independent constraints.

\subsection{Consistency conditions}

So far we have identified the first class constraints generating all the known gauge symmetries  listed in Section {\bf \ref{Sec:Lagr}} except the time diffeomorphisms. Moreover, we need to understand the conditions on the Lagrange multipliers that arise from the demand that the constraints be preserved in time (``consistency conditions''):
\begin{equation}
{d {\cal G}^{\alpha}_{\ \beta}  \over dt}=0 \, , \quad {d\phi^{i\alpha}_{\ \ \beta } \over dt } = 0\, , \quad {d\pi^{\kappa\lambda} \over dt } = 0 \, .
\end{equation}
These two problems are  related.  
The consistency algorithm will in fact show that there are yet three hidden gauge symmetries to be uncovered, including the searched-for time diffeomorphisms. 

On-shell time derivatives are computed by taking Poisson brackets with the Hamiltonian,
\begin{equation}
H = \int d^3 x (\Gamma^{\beta}_{\ \alpha 0 }\, {\cal G}^{\alpha}_{\ \beta}  +X^{\alpha}_{\ \beta\, i} \phi^{i \beta}_{\ \alpha}  +Y_{\alpha\beta} \pi^{\alpha\beta})
\end{equation}
This is easy since we know all Poisson brackets among the constraints. 

We start with the easiest, ${d {\cal G}^{\alpha}_{\ \beta}  \over dt}=0$. Since this constraint commutes weakly with itself and all others, it is automatically preserved in time. No conditions arise from this consistency condition. The corresponding Lagrange multipliers $\Gamma^{\alpha}_{\ \beta0}$ are therefore left fully free.  This is as it should be since ${\cal G}^{\alpha}_{\ \beta}$ have been identified to be the generators of the $GL(4)$ gauge symmetry.

The other constraints, which exhibit nice geometrical properties of the action (\ref{GB}), need a more detailed analysis. Using the Poisson bracket relations the consistency equations are
\begin{eqnarray}
{d\phi^{i\alpha}_{\ \ \beta } \over dt } &=& \Omega^{\kappa\lambda\,\alpha\gamma}_{\ \ \ \ \  \beta\delta}\,\epsilon^{ijk}\Big(-2 \nabla_{j} g_{\kappa\lambda}\,  X^{\delta}_{\ \gamma k } +  R^{\delta}_{\ \gamma\, jk} Y_{\kappa\lambda}\Big) =0 \label{cons1} \\
{d\pi^{\kappa\lambda} \over dt } &=& \Omega^{\kappa\lambda\, \alpha\gamma}_{\ \ \ \ \ \beta\delta} \,\epsilon^{ijk}\,  R^{\beta}_{\ \alpha\, ij} X^{\delta}_{\ \gamma\, k } =0 \label{cons2}
\end{eqnarray} 
The unknowns in these equations are the Lagrange multipliers $X^{\gamma}_{\  \delta i}$ and $Y_{\alpha\beta}$. 

The consistency equations (\ref{cons1}) and  (\ref{cons2}) are exactly equivalent to (\ref{3p12}) and (\ref{3p11}) via the replacements
$X^{\alpha}_{\  \beta i } = R^{\alpha}_{\ \beta 0i  } $ and 
$Y_{\alpha\beta} = \nabla_0 g_{\alpha\beta}$,
which we have derived above (Eqs (\ref{Eq:Y}) and  (\ref{Eq:X})).

There are no further consistency conditions to be checked once the Lagrange multipliers  $X^{\gamma}_{\  \delta i}$ and $Y_{\alpha\beta}$ are known fulfilling (\ref{cons1}) and  (\ref{cons2}).  This last step (understanding the freedom in $X$ and $Y$), still to be achieved, will end therefore the consistency algorithm.

\section{Counting the number of degrees of freedom (4D)}
\label{Sec:4DCounting}

\subsection{The equations}

The naive number of equations for $X^{\gamma}_{\  \delta i}$ and $Y_{\alpha\beta}$ coincide with the number of unknowns. However, due to the symmetries there are less independent equations. 

With the geometrical insight provided by the invariance of the action, we can write at once a family of solutions to (\ref{cons1}) and (\ref{cons2}), 
\begin{eqnarray}
X^{\alpha}_{\  \beta i} &=& R^{\alpha}_{\ \beta\, ij }\, \xi^j + w_i \delta^\alpha_\beta  \label{sol1}  \\
Y_{\alpha\beta} &=& \rho\,  g_{\alpha\beta} +  \xi^i \nabla_i g_{\alpha\beta}. \label{sol2}
\end{eqnarray}
where the parameters $\xi^i, w_i$ and $\rho$ implement diffeomorphisms, projective and Weyl transformations, respectively. These parameters span a vector space of dimension $3+3+1 = 7$.

To explore whether there are other solutions to  (\ref{cons1}) and (\ref{cons2}), which would imply further ambiguity in the solutions and hence further gauge symmetries, we write this system in matrix form, 
\begin{equation}
\left(  \begin{array}{cc}
\{\phi^{i\alpha}_{\ \ \beta},\phi^{j\gamma}_{\ \ \delta}\} & \{\phi^{i\alpha}_{\ \ \beta},\pi^{\kappa\lambda}\}  \\ 
\{\pi^{\delta\gamma},\phi^{j\gamma}_{\ \ \delta}\} & 0
\end{array}   \right)  \left( \begin{array}{c}
X^{\delta } _{\ \gamma j} \\ 
Y_{\kappa\lambda}
\end{array}  \right) =0 \ \Leftrightarrow \ \mathbf{M}\left( \begin{array}{c}
X^{\delta } _{\ \gamma j} \\ 
Y_{\kappa\lambda}
\end{array}  \right) = 0 \, , \label{meq}
\end{equation}
where
\begin{equation}
\mathbf{M} = \left( \begin{array}{cc}
P & Q \\ 
-Q^t & 0
\end{array}   \right) \, ,
\end{equation} 
and where the matrices $P$ and $Q$ have matrix elements
\begin{eqnarray}
\{\phi^{i\alpha}_{\ \ \beta},\phi^{j\gamma}_{\ \ \delta}\} &=& \Omega^{\kappa\lambda\alpha\gamma}_{\ \ \ \ \ \beta\delta}\epsilon^{ijk}\nabla_{k}g_{\kappa\lambda}  \\
\{\phi^{i\alpha}_{\ \ \beta},\pi^{\kappa\lambda}\}   &=&  \Omega^{\kappa\lambda\alpha\gamma}_{\ \ \ \ \ \beta\delta} \epsilon^{ijk} R^{\delta}_{\ \gamma jk} \, .
\end{eqnarray} 
We must study the rank of the $58 \times 58$ matrix $\mathbf{M}$ in order to determine the number of parameters the linear system (\ref{cons1}) and (\ref{cons2}) leaves free. 

As we pointed out earlier, the rank of this matrix depends on the location in phase space.  For instance, if both $R^{\beta}_{\ \alpha ij}$ and $\nabla_{k}g_{\kappa\lambda}$ vanish, the matrix degenerates to zero.  We are interested in the generic value of the rank, valid for generic points on the constraint surface, which is the only restriction to be imposed on $R^{\beta}_{\ \alpha ij}$ and $\nabla_{k}g_{\kappa\lambda}$.  This restriction was determined above to be
\begin{equation}
\Omega^{\kappa\lambda\alpha\gamma}_{\ \ \ \ \ \beta\delta } \epsilon^{ijk} R^{\beta}_{\ \beta\, i j}\nabla_k g_{\kappa\lambda}=0  .\label{cfin}
\end{equation}

\subsection{Strategy for analyzing the constraint on $R^{\beta}_{\ \alpha ij}$ and $\nabla_{k}g_{\kappa\lambda}$}

The constraint (\ref{cfin}) is simple to solve in the vicinity of a given point $x^\alpha$, as we now show. 

The free indices in (\ref{cfin}) are $\gamma, \delta$, but the trace is identically zero. These are then 15 equations for the 10 components of the metric $g_{\alpha\beta}$ plus 48 components of the (spatial) connection $\Gamma^{\alpha}_{\ \beta i }$.  

The key property is multi-linearity. Due to the Levi-Civita symbol $\epsilon^{ijk} $, any component, say $\Gamma^{\alpha}_{\ \beta\, 3 }$, never appears multiplied by itself.  

The constraint is also invariant under local  GL(4) transformations. This means without loss of generality we can always set 
\begin{eqnarray}
g_{\alpha\beta} = \eta_{\alpha\beta},
\end{eqnarray} 
and the constraint becomes an equation for the connection:
\begin{eqnarray}\label{c00}
2\Omega^{\kappa\lambda\alpha\gamma}_{\ \ \ \ \ \beta\delta } \epsilon^{ijk} R^{\beta}_{\ \alpha\, i j} \Gamma_{\kappa\lambda\, k}=0 .
\end{eqnarray}

Spatial translational invariance allows us to sit at the point $\{x,y,z\} = \{0,0,0\}$.   The curvature depends on first derivatives of the connection so it is enough to assume a first order expansion, 
\begin{equation}\label{c}
\Gamma^{\alpha}_{\  \beta \, i}(x) = A^{\alpha}_{\  \beta \, i} + B^{\alpha}_{\  \beta \, i j}\, x^j
\end{equation}
 where $A, B$ are yet unknown matrices. 

~ 
 
 The algorithm is the following:
 
 \begin{enumerate}
 \item  Replace the connection (\ref{c}) in (\ref{c00}) . After all derivatives have been computed set 
 \begin{eqnarray}
 \{x,y,z\} = \{0,0,0\}
 \end{eqnarray} 
 \item The constraint becomes a multilinear function of all the coefficients $A^{\alpha}_{\  \beta \, i},\, B^{\alpha}_{\  \beta \, ij}$. These are 
\begin{eqnarray}
16 \times 3 \ + \ 16 \times 9 = 192
\end{eqnarray}  
 coefficients. 
  
 \item  Give random values to 192 - 15  = 177 of them and solve the constraint for the 15 free values.  For example, give random values for
\begin{eqnarray}
B^{\alpha}_{\  \beta \, ij},\,  A^{\alpha}_{\  \beta \, 1}, \,  A^{\alpha}_{\  \beta \, 2}
\end{eqnarray}  
leaving $ A^{\alpha}_{\  \beta \, 3}$ as the unknowns.  Thanks to multilinearity, the equations (\ref{c00}) become a system of 15 independent linear equations for the 16 $ A^{\alpha}_{\  \beta \, 3}$'s.  It is trivially solved. One coefficient (and only one) remains free, which can be chosen at random or left free. 
 
 \item  Replace all these numbers in the matrix $M$ and compute the rank.   
 \end{enumerate}
 
\subsection{Results and number of degrees of freedom}
Since the algorithm generates solutions of the constraint (\ref{cfin}) chosen at random, it provides the generic value of the rank.

The algorithm has been run countless times.  
One always  finds
\begin{equation}
\mbox{Rank}(\mathbf{M}) = 48 \, .
\end{equation}
We note also
\begin{eqnarray}
\mbox{Rank}(P) &=& 36 \\
\mbox{Rank}(Q) &=& 9 \, .
\end{eqnarray}
Substracting the rank of the matrix $\mathbf{M}$ from its dimension, we conclude that it possesses $10$ zero vectors.

Now, the known zero eigenvectors of the matrix $\mathbf{M}$, corresponding to the already identified gauge symmetries (spatial diffeomorphisms, spatial projective transformations and Weyl transformations) span a vector space of $7$ dimensions.  Accordingly, there are $10-7=3$ additional zero eigenvectors.

Let $X^{\gamma}_{A\  \delta i}$ and $Y_{A \,  \gamma\delta}$ ($A=1, 2, 3$) be three independent solutions of the linear system (\ref{meq}), which are also independent from the $7$ known solutions, so that the general solution reads
\begin{eqnarray}
X^{\alpha}_{\  \beta i} &=& R^{\alpha}_{\ \beta\, ij }\, \xi^j + w_i \delta^\alpha_\beta +\alpha^A X^{\alpha}_{A\  \beta i}  \label{sol12}  \\
Y_{\alpha\beta} &=& \rho\,  g_{\alpha\beta} +  \xi^i \nabla_i g_{\alpha\beta} + \alpha^A Y_{A \,  \alpha\beta}  \, , \label{sol22}
\end{eqnarray}
where $\alpha^A$ are arbitrary. Since the zero eigenvectors define first class constraints, the solutions $X^{\alpha}_{A\  \beta i}$ and $Y_{A \,  \alpha\beta}$ define  $3$ additional gauge symmetries, which are again somewhat hidden, in the sense that they are not present in the list of Subsection {\bf \ref{Subsec:GaugeSym}} of ``obvious" gauge symmetries. Because $X^{\alpha}_{A\  \beta i}$ and $Y_{A \,  \alpha\beta}$ are phase space functions, the new symmetries are expressed in terms of the canonical variables. A first hint of extra symmetries can be  seen when solving the equation of motion with spherical symmetry. See  \cite{Banados:2025uhb} for details on the torsionless case.  

One could write in principle the null vectors $X^{\alpha}_{A\  \beta i}$ and $Y_{A \,  \alpha\beta}$ of the matrix $\mathbf{M}$ in terms of its matrix elements, which are phase space functions. Although additional insight can be provided on the new symmetries (see next subsection), we have not been able, however, to get manifestly covariant expressions for them after a few simple tries.  This is somewhat curious and unexpected but other systems are known to exhibit similar features, one prime example being the superparticle  \cite{Casalbuoni:1975hx,Casalbuoni:1976tz,Brink:1981nb,Brink:1981rt,Bengtsson:1984rw,Brink:1987bi}.

On general grounds, an odd number of extra gauge symmetries was actually expected. The reason is that the rank of the antisymmetric matrix $\mathbf{M}$ should be even, which would not be the case if there were only $7$ zero eingenvectors.   

The rank of $Q$ is perhaps not surprising. It is a 48$\times$10 matrix and the Weyl transformation, which acts only on the metric, $Y_{\alpha\beta}=2\delta g_{\alpha\beta}$, is a known solution.

Knowing the rank of $\mathbf{M}$, one can count the number of physical degrees of freedom.  Denoting by $X$ the \textbf{three} new gauge symmetries, 
we have the following table: 

\vspace{.3cm}

\begin{tabular}{|l|l|r|}
\hline 
Fields & $\{p^{i\alpha}_{\ \beta },\ A^{\gamma}_{\ \delta j}\}, \ \ \{g_{\alpha\beta},\ \pi^{\gamma\delta}\} $ & $2\times 3\times 16\ +\ 2\times 10 = 116$ \\ 
\hline 
Constraints & ${\cal G}^{\alpha}_{\ \beta}, \ \phi^{i\gamma}_{\ \delta }, \ \pi^{\kappa\lambda}  $ & $15  + 3\times 16 +  10 =~ 73$   \\ 
\hline 
Symmetries & gl(4), diff, projective, Weyl,  \textbf{X} & $16 + 3  + 3+1 -1 + \mathbf{3} =~ 25$ \\ 
\hline 
Second class &  & $73-25=48$ \\
\hline
 Physical dof & fields-2$\times$(first)-(second) & $116 - 2\times 25 -48 = 18$  \\ 
\hline 
\end{tabular} 

\vspace{.3cm}

The $-1$ in the counting of the  symmetries takes care of the reducibility relation.

This theory then has $9$ physical degrees of freedom.

\subsection{Some insight on the extra gauge symmetries}

We now come back to the extra solutions of the equations  (\ref{meq}).  The random algorithm providing a matrix $\mathbf{M}$ with maximal rank also gives us two interesting pieces of information:

\begin{enumerate}
\item  First, the extra three vectors may be taken to have $Y_{\alpha\beta}=0$ and therefore $X^{\alpha}_{\ \beta \, i }$ satisfies, simultaneously, 
  \begin{eqnarray}
\Omega^{\kappa\lambda\alpha\gamma}_{\ \ \ \ \ \beta\delta}\epsilon^{ijk}\nabla_{k}g_{\kappa\lambda}\, X^{\delta}_{\ \gamma\,j } &=&0, \label{eq1} \\
 \Omega^{\kappa\lambda\alpha\gamma}_{\ \ \ \ \ \beta\delta} \epsilon^{ijk} R^{\delta}_{\ \gamma jk} X^{\beta}_{\ \alpha\,i } &=& 0. \label{eq2}
\end{eqnarray} 
The projective symmetry is the solution $X^{\alpha}_{\ \beta\,i} = w_{i}\delta^\alpha_\beta$ ($w_i$ arbitrary), already counted. The random matrices $\mathbf{M}$ produced by the algorithm show that there exists three extra solutions to (\ref{eq1}) and (\ref{eq2}).

\item  Second,  solutions to (\ref{eq1}) and (\ref{eq2}) have the property, 
\begin{eqnarray}
X^{\alpha\beta}_{\ \ \ i}=X^{\beta\alpha}_{\ \ \ i} 
\end{eqnarray} 
where we have raised the index with the metric. This is not at all obvious from (\ref{eq1}) and (\ref{eq2}) but nevertheless true.
\end{enumerate}

Translating this information to the variation of fields we conclude that the three missing ``$X$" symmetries have the properties:
\begin{enumerate}
\item Since $Y_{\alpha\beta}=0$, they do not act on the metric, 
\begin{eqnarray}
\delta g_{\alpha\beta}=0. 
\end{eqnarray} 

\item  Since $X^{\alpha\beta}_{\ \ \ i}$ is symmetric, the variations of the connection satisfy
\begin{eqnarray}
\delta \Gamma^{\alpha\beta}_{\ \ \ i}=\delta \Gamma^{\beta\alpha}_{\ \ \ i}. 
\end{eqnarray} 
\end{enumerate}

Motivated by the above comments, let us go back to the Lagrangian and look for symmetries such that, 
\begin{eqnarray}
\delta g_{\alpha\beta} &=& 0, \label{v1}\\
\delta \Gamma^{\alpha\beta}_{\ \ \ \mu} &=& \delta \Gamma^{\beta\alpha}_{\ \ \ \mu}.  \label{v2}
\end{eqnarray} 
We will be able to find {\it two} symmetries of the Lagrangian satisfying these properties,  on top of 4 projective transformations, which are also of this form.

In order to identify these symmetries,   we use from now on differential form notations hiding the spacetime index. Varying the Gauss-Bonnet action with (\ref{v1}) and (\ref{v2}) we have 
\begin{eqnarray}
\delta I = \int \sqrt{g}\epsilon_{\alpha\beta\gamma}^{\ \ \ \ \, \delta} R^{\alpha\beta}\wedge \nabla \delta\Gamma^{\gamma}_{\ \delta}  \ \ \ \ \ \ \ \ \   (\delta R^{\alpha}_{\ \beta}= \nabla \delta\Gamma^{\alpha}_{\ \beta} ) \nonumber
\end{eqnarray} 
Without any loss of generality, we expand the connection 1-form $\Gamma^{\alpha}_{\ \beta}$ as,
\begin{eqnarray}
\Gamma^{\alpha}_{\ \beta} = \overset{(g)}{\Gamma}~\!\!^{\alpha}_{\ \beta} + A^{\alpha}_{\ \beta} + S^{\alpha}_{\ \beta}
\end{eqnarray} 
where $\overset{(g)}{\Gamma}~\!\!\!^{\alpha}_{\ \beta}$ is the Levi-Civita part, and we have split the remaining
into symmetric and antisymmetric parts, that is, 
\begin{eqnarray}
\overset{(g)}{\nabla}_{\alpha}\,  g_{\gamma\delta} =0, \ \ \ \  A^{\alpha\beta}= - A^{\beta\alpha},  \ \ \  S^{\alpha\beta}= S^{\beta\alpha}. 
\end{eqnarray} 

A short calculation shows that the action variation becomes (wedge symbols omitted), 
\begin{eqnarray}
\delta I = \int \sqrt{g} \epsilon_{\alpha\beta\gamma\delta} R^{\alpha\beta}\Big( \underbrace{\overset{(g)}{\nabla} \delta\Gamma^{\gamma\delta}+ A^{\gamma}_{\ \sigma}\delta\Gamma^{\sigma\delta}+A^{\delta}_{\ \sigma}\delta\Gamma^{\sigma\gamma}}_{\mbox{symmetric($\gamma,\delta$)}} + \underbrace{S^{\gamma}_{\ \sigma}\delta\Gamma^{\sigma\delta}-S^{\delta}_{\ \sigma}\delta\Gamma^{\sigma\gamma}}_{\mbox{antisymmetric($\gamma,\delta$)}} \Big)   \nonumber
\end{eqnarray}  
Note that this splitting is valid assuming the connection variation satisfies (\ref{v2}).

The first bracket, symmetric in $\gamma,\delta$ vanishes against $\epsilon_{\alpha\beta\gamma\delta}$. We conclude that symmetric variations $ \delta \Gamma^{\alpha\beta}$ satisfying the algebraic relation, 
\begin{eqnarray}
 S^{\alpha}_{\ \gamma}\wedge \delta\Gamma^{\gamma\beta}- S^{\beta}_{\ \gamma}\wedge \delta\Gamma^{\gamma\alpha}  =0 \label{cond}
\end{eqnarray} 
define gauge symmetries of the action. This is an equation for $\delta\Gamma^{\sigma\gamma} $,  we now study its solutions.

Projective solutions once again  show up here, $\delta\Gamma^{\sigma}_{\ \gamma} = w\, \delta^{\alpha}_{\beta}$ ($w$= arbitrary 1-form) are four solutions. We already know these ones. 

The important comment is that there exists {\it two} other solutions. We have found these by giving random values to $S^{\alpha}_{\ \beta}$ and solve the linear problem (\ref{cond}) for $\delta\Gamma^{\sigma\gamma}$. This process always give $4\mbox{(projective)}\ +\ 2$ solutions.  

Concluding, the symmetric transformations (\ref{v2}) explain two of the three hidden symmetries. 

The last hidden symmetry does have the property $\delta \Gamma^{\alpha\beta}_{\ \ \ i } = \delta \Gamma^{\beta\alpha}_{\ \ \ i}$ for  spatial indices. We know this from the above Hamiltonian derivation. Since the present analysis indicates that the stronger condition (\ref{v2}) only covers 2 symmetries, the 3rd symmetry does not fulfill the ansatz (\ref{v2}), i.e.,  has the property that $\delta \Gamma^{\alpha\beta}_{\ \ \ 0 } \neq \delta \Gamma^{\beta\alpha}_{\ \ \ 0}$.  This renders the task of writing it in a covariant way difficult.

We close by some comments on the time diffeomorphisms, which read, in their improved form 
\begin{eqnarray}
\delta \Gamma^{ \alpha}_{\ \beta i} &=& R^{\alpha}_{\ \beta i 0}\, \xi^0 \nonumber \\
\delta g_{\alpha\beta}&=& \nabla_0 g_{\alpha\beta} \, \xi^0 \label{tdiff}
\end{eqnarray}
Contrary to the time component of the projective transformation, which acts only on the Lagrange parameters and are irrelevant for the phase space variables, the transformations  (\ref{tdiff}) do have a non-trivial phase space action.

Since  $ R^{\alpha}_{\ \beta\gamma 0}$ and $\nabla_0 g_{\kappa\lambda} $ are equal to $X^{\alpha}_{\ \beta i }$ and $Y_{\alpha\beta}$ (see (\ref{Eq:X}),(\ref{Eq:Y})), we have according to (\ref{sol12}) and  (\ref{sol22}),
\begin{eqnarray}
R^{\alpha}_{\ \beta 0i} &=& R^{\alpha}_{\ \beta i j} \zeta^j + v_i \delta^{\alpha}_{\ \beta}+\beta^A X^{\gamma}_{A\  \delta i} \, , \label{t2}\\
\nabla_0 g_{\alpha\beta} &=& \sigma g_{\alpha\beta} + \zeta^i \nabla_i g_{\alpha\beta} + \beta^A Y_{A \,  \gamma\delta}  \label{t1} 
\end{eqnarray} 
for some $(\zeta^i, v_i, \sigma, \beta^A)$. If the $\beta^A$ associated with ``the last"  of the extra gauge symmetries is different from zero, the time diffeomorphisms do involve it and one could express it in terms of time diffeomorphisms.   One could thus say  that time diffeomorphisms provide ``the 3rd"  extra gauge symmetry.  However, this characterization depends on the choice made for the Lagrange multipliers, which occur in $R^{\alpha}_{\ \beta 0i} $ and $\nabla_0 g_{\alpha\beta}$.  For that reason,  it would not be entirely satisfactory -- a phenomenon we already encountered in $2$ spacetime dimensions.

\section{Conclusions}
\label{Sec:Conclusions}

In this paper, we have studied the dynamics of the Palatini Gauss-Bonnet theories defined by the action (\ref{GB0}).  We have worked out the Hamiltonian formulation in $2$ and $4$ spacetime dimensions.  The analysis is direct in $2$ dimensions but more intricate in $4$ dimensions, where there are non trivial degrees of freedom.  These degrees of freedom are less easy to identify because of the nonlinearities.  The same features hold in Chern-Simons theories in $2n+1$ dimensions.

A central property of the models is that they possess additional gauge symmetries besides the ones that one can easily spot by direct inspection of the Lagrangian. This is perhaps not too surprising given the close parenthood of the models with topological theories.  However, these symmetries are rather intrincate to write down in $D \geq 4$ dimensions  and were revealed through the Hamiltonian analysis.  One of their curious features is that they do not have an obvious manifestly covariant expression in $4$ dimensions.

An unusual aspect of our approach is the use of a random algorithm for determining the maximum possible rank of the matrix relevant for counting the gauge symmetries.  The idea is that random draws will hit the generic region where this rank takes its maximum  value and not the zero measure submanifolds where the rank is lower.   This algorithm provides furthermore interesting information on the form of the gauge symmetries. 

We reserve for future work the analysis of boundary conditions and boundary dynamics, which is relevant when there are boundaries or asymptotic regions.

\section*{Acknowledgements}
 The work of MH is partially supported by FNRS-Belgium (convention IISN 4.4503.15), as well as by research funds from the Solvay Family.  MB would like to thank S. Theisen, R. Emparan and L. Garay for useful conversations and hospitality in Golm, Barcelona, Madrid, respectively, and ULB and the International Solvay Institute for hospitality in Brussels. He would also like to thank J.M. Martín-García, L. Stein, J. Margalef and A. García-Parrado for useful advice on the xAct [https://xact.es/] system.  M.B was partially funded by a Grant from Vicerrectoría Académica UC-Chile.

\providecommand{\href}[2]{#2}\begingroup\raggedright\endgroup

\end{document}